# Origins of *de novo* genes in human and chimpanzee

Jorge Ruiz-Orera[1], Jessica Hernandez-Rodriguez[2], Cristina Chiva[3,4], Eduard Sabidó[3,4], Ivanela Kondova[5], Ronald Bontrop[5], Tomàs Marqués-Bonet [2,6,7], M.Mar Albà[1,2,7,]*

[1]Evolutionary Genomics Group, Hospital del Mar Research Institute (IMIM), Dr. Aiguader 88, Barcelona 08003, Spain

[2]Department of Experimental and Health Sciences, Universitat Pompeu Fabra (UPF), Dr. Aiguader 88, Barcelona 08003,Spain

[3]Proteomics Unit, Universitat Pompeu Fabra (UPF), Dr. Aiguader 88, Barcelona 08003, Spain

[4]Proteomics Unit, Centre de Regulació Genòmica (CRG), Dr. Aiguader 88, Barcelona 08003, Spain

[5]Biomedical Primate Research Center (BPRC), Lange Kleiweg 161, 2288 GJ Rijswijk, The Netherlands

[6]Centro Nacional de Análisis Genómico (CNAG), Baldiri Reixac 4, Barcelona 08028, Spain

[7]Institució Catalana de Recerca i Estudis Avançats (ICREA), Pg. Lluis Companys 23, Barcelona 08010, Spain

*Corresponding author: M.Mar Albà (malba@imim.es)

**Abstract**

The birth of new genes is an important motor of evolutionary innovation. Whereas many new genes arise by gene duplication, others originate at genomic regions that do not contain any gene or gene copy. Some of these newly expressed genes may acquire coding or non-coding functions and be preserved by natural selection. However, it is yet unclear which is the prevalence and underlying mechanisms of *de novo* gene emergence. In order to obtain a comprehensive view of this process we have performed in-depth sequencing of the transcriptomes of four mammalian species, human, chimpanzee, macaque and mouse, and subsequently compared the assembled transcripts and the corresponding syntenic genomic regions. This has resulted in the identification of over five thousand new transcriptional multiexonic events in human and/or chimpanzee that are not observed in the rest of species. By comparative genomics we show that the expression of these transcripts is associated with the gain of regulatory motifs upstream of the transcription start site (TSS) and of U1 snRNP sites downstream of the TSS. We also find that the coding potential of the new genes is higher than expected by chance, consistent with the presence of protein-coding genes in the dataset. Using available human tissue proteomics and ribosome profiling data we identify several *de novo* genes with translation evidence. These genes show significant purifying selection signatures, indicating that they are probably functional. Taken together, the data supports a model in which frequently-occurring new transcriptional events in the genome provide the raw material for the evolution of new proteins.

**Author summary**

The presence in every genome of a set of genes which are unique to the species, lacking homologues in other species, has puzzled scientists for the past 20 years. How have these genes originated? The advent of massively parallel RNA sequencing (RNA-Seq) has provided new clues, with the discovery of an unsuspectedly high number of transcripts that do not correspond to typical protein-coding genes and which could serve as a substrate for this process. Here we have examined RNA-Seq data from several mammalian species in order to define a set of putative newborn genes in human and chimpanzee and investigate what drives their expression. This is the largest-scale project that tries to address this question. We have found thousands of transcripts that are human- and/or chimpanzee-specific and which are likely to have originated *de novo* from previously non-transcribed regions of the genome. We have observed an enrichment in transcription factor binding sites in the promoter regions of these genes when compared to other species, consistent with the idea that the gain of new regulatory motifs results in *de novo* gene expression. We also show that some of the genes encode new functional proteins expressed in brain or testis, which may have contributed to phenotypic novelties in human evolution.

**INTRODUCTION**

New genes continuously arise in genomes. Evolutionary 'inventions' include small proteins that have functions related to the adaptation to the environment, such as antimicrobial peptides or antifreeze proteins, which have independently evolved in different groups of organisms [1,2]. Whereas many genes arise by gene duplication [3], other evolve *de novo* from previously non-genic regions in the genome [4–10]. These genes are only detected in one or a few related species [4,5,11]. Syntenic genomic regions from more distant species do not express any similar genes. In *de novo* gene evolution the full sequence space can be potentially explored for new adaptive functions, without the limitations imposed by high similarity to an already existing gene [12]. In *Drosophila*, recently emerged genes are often essential [13,14], highlighting their functional importance.

Recent efforts to characterize *de novo* originated genes in humans and other primates have focused on annotated protein-coding genes [5,15–17]. For example Chen and colleagues have identified 64 human protein-coding genes that correspond to putative long non-coding RNAs (lncRNAs) in macaque and proposed that these human proteins have originated from previously existing non-coding transcripts [17]. Other studies have focused on specific genes. One case is the hominoid-specific antisense gene NCYM, which is over-expressed in neuroblastoma [18]. This gene inhibits the activity of glycogen synthase kinase 3β (GSK3β), which targets NMYC for degradation. Another example is FLJ33706, which encodes a 194 amino acid protein expressed in different human brain structures [19]. This type of genes, sometimes called orphans, have been described in a wide range of organisms, including yeast [20,21], mouse [22], rat [23], insects [24–26] and plants [27,28]. They share common features, such as short protein size and tissue-specific expression [6,29].

Massively parallel RNA sequencing (RNA-Seq) has revealed that a large fraction of the genome, extending far beyond the set of annotated genes, is transcribed [30,31] and possibly translated [32–36]. Many genes that are annotated as lncRNAs are lineage-specific and display high transcriptional turnover [37,38]. These observations have important implications for *de novo* gene evolution, as it means that there is abundant raw material for the evolution of new functional proteins.

Here we use transcriptomics data from four mammalian species to quantify the amount of transcription that is human- and/or chimpanzee-specific and investigate the molecular mechanisms driving the expression of these genes. The majority of *de novo* genes originate from regions with conserved genomic synteny in macaque. Analysis of these regions reveals that the expression of the gene is associated with the gain of novel regulatory motifs in the promoter region and U1snRNP splice sites downstream of the transcription start site. We also show that at least a subset of the newly evolved genes are likely to encode functional proteins.

**RESULTS**

**Assembly of annotated and novel transcripts from strand-specific RNA-Seq data**

We used strand-specific sequencing of polyadenylated RNA (polyA+ RNA-Seq) from several tissues from human, chimpanzee, macaque and mouse, to perform *de novo* transcript assembly with Cufflinks [39]. The total number of RNA-Seq datasets was 43, of which 26 were generated in this study and the rest were public datasets from previous studies [37,40,41]. The set of tissues sampled included testis and brain, previously noted to be enriched in *de novo* genes [16,40]. In this study, we will refer to gene as the set of transcripts merged into a single loci by Cufflinks. Any genome unmapped reads were assembled with Trinity for the sake of completeness [42].

Subsequently we selected transcripts with a size longer than 300 nucleotides (nt). This removed any sequencing products resulting from one single amplified paired end read (2x100 nt). We also filtered out all those genes having a per-base read coverage lower than 5, to ensure transcript completeness (see Materials and Methods). A negative control lacking reverse transcriptase in the library construction step (RT-) indicated that the probability of a transcript to represent DNA contamination was very low, virtually 0 in the case of multiexonic transcripts. To ensure a highly robust set of transcripts we filtered out intronless genes. This also removed possible promotor- or enhancer-associated transcripts (PROMPTS and eRNAs). As a result of this process, we recovered 99,670 human, 102,262 chimpanzee, 93,860 macaque and 85,688 mouse transcripts merged in 34,188 human, 35,915 chimpanzee, 34,427 macaque and 31,043 mouse gene loci. This included a high fraction of the long multiexonic genes annotated in each species plus a significant number of additional non-annotated genes (Figure 1A). The number of non-annotated genes was much larger in chimpanzee and macaque than in human and mouse, mostly due to the inclusion of many lncRNAs in the human and mouse annotations. Not surprisingly, novel genes tended to be shorter and expressed at lower levels than annotated genes (Figures 1B and 1C). In humans, annotated genes represented about 98% (testis) and 99-99.5% (rest of tissues) of the transcriptional cost as measured in terms of sequencing reads.

**Identification of *de novo* human and chimpanzee genes**

Next, we used BLAST-based sequence similarity searches [43] to identify the subset of *de novo* genes that could have originated in human, chimpanzee, or the common ancestor of these two species since the divergence from macaque (hominoid-specific genes). These genes survived an exhaustive search for homologues in related species, including the transcript assemblies described above, the transcript assemblies we obtained using a similar procedure but using previously published non-stranded single read RNA-Seq data for nine vertebrate species [44],

Ensembl gene annotations for the same set of species, and the complete expressed sequence tag (EST) and non-redundant (nr) protein databases from NCBI. We also employed genomic alignments to discard any transcripts expressed in syntenic regions in other species that could have been missed by BLAST (S1 Fig). This pipeline resulted in 634 human-specific genes (1,029 transcripts), 780 chimpanzee-specific genes (1,307 transcripts) and 1,300 hominoid-specific genes (3,062 transcripts). By adding these numbers the total number of candidate *de novo* genes was 2,714 (5,398 transcripts) (Figure 2A).

As we used strand-specific RNA sequencing we could unambiguously identify a large number of antisense transcripts. Many of them were located within intronic regions (38.31%) and others partially overlapped exonic regions from other genes (10.62%). The rest of *de novo* transcripts were located in ntergenic regions (51.07%). These percentages were similar for human-, chimpanzee- and hominoid-specific genes (Figure 2B). Eight *de novo* genes matched annotated protein-coding genes (S1 Table). One example was the gene *GTSCR1* (Gilles de la Tourette syndrome chromosome region, candidate 1), encoding a 137 amino acid long protein and with proteomics evidence for 23.53% of the sequence. Curiously, the human protein-coding genes, including GTSCR1, were annotated as long non-coding RNAs (lncRNAs) in a subsequent Ensembl version (77). About 20% of *de novo* genes matched annotated lncRNAs or sequence entries in the 'EST' or 'nr' databases. The identification of lncRNAs in the set was not surprising as it has been previously reported that they tend to show low phylogenetic conservation [37,38,45]. The majority of *de novo* genes were novel genes (~ 80%, Figure 2C).

Transcripts from *de novo* genes were shorter and expressed at lower levels than those from conserved annotated genes (S2 Fig). These biases have also been noted in young annotated primate protein-coding genes [5,40]. In general, *de novo* genes were located in regions with conserved synteny in macaque (> 75%, S3 Fig). The proportion of *de novo* transcripts with complete synteny was similar to the one observed for conserved transcripts. *De novo* transcripts were enriched in transposable elements, about 20% of their total transcript length was covered by transposable elements compared to only about 8% for conserved genes (S4 Fig). An enrichment in transposable elements was previously observed in primate-specific protein-coding genes [5] and in lncRNAs in general [46].

**D*e novo* genes are enriched in testis**

We determined the number of human and chimpanzee genes expressed in different tissues using the RNA-Seq data. The vast majority of *de novo* transcripts were expressed in testis (93.8-94.5%), as were transcripts that showed higher conservation levels (Figure 2D). In contrast, in brain, liver and heart, *de novo* transcripts were underrepresented when compared to other transcripts. The gene expression enrichment in testis resembles the bias observed for mammalian lncRNAs [41,47,48]. The high number of *de novo* genes detected in testis did not appear to be the result of increased capacity to detect weakly expressed genes in this tissue as deduced from the overall distribution of gene expression values (S5 Fig). It was previously reported that young human protein-coding genes were enriched in the brain [16], but we did not detect a similar bias in our data.

As a result of the expression tissue differences mentioned earlier, *de novo* genes were twice as likely to show testis restricted expression as the rest of genes (94.1%-96.4% in contrast to ~64% of all assembled transcripts, see Material and Methods). The use of gene expression data from GTEx, although restricted to human annotated transcripts, produced consistent results (S6 Fig). The majority of *de novo* genes were detected in all or nearly all the 60 individuals tested for testis in GTEx [49], indicating that they are expressed in a stable manner in the population (Figure 2E).

**Signatures of transcription initiation and elongation in *de novo* genes**

Divergent transcription from bidirectional promoters is widespread in eukaryotic genomes [50,51] and leads to the expression of numerous antisense transcripts, most of them poorly conserved in other species and generally lacking coding potential [52]. It has been proposed that the reuse of existing promoters can be a driving force of new gene origination [53]. We searched for bidirectional promoters by scanning the genome for transcription start sites of antisense transcripts at a distance < 1 Kb. Positive cases showed a typical separation between the two TSSs of about 100 bp, consistent with the presence of a bidirectional promoter (S7 Fig) However, *de novo* genes were not enriched in bidirectional promoters with respect to the rest of genes (20% versus 29.81%), indicating that this is probably not the predominant mechanism for the formation of *de novo* genes.

Comparison of GC content in the region surrounding the TSSs clearly revealed that *de novo* genes are more A/T-rich than conserved annotated genes (S8 Fig). We searched for overrepresented transcription factor binding sites in the promoters of *de novo* genes using the programs PEAKS [54] and HOMER [55] (Figure 3A and 3B). With PEAKS we identified a strong enrichment of sites for CREBP, RFX, and JUN in the first 100 bp upstream of the TSS (p-value < $10^{-5}$, motif frequency > 20% higher than in other regions). Whereas CREBP (cAMP-responsive element binding protein) and JUN (transcription factor AP1) are general activators, RFX (regulatory factor X) has been associated with expression in testis [56,57]. With HOMER we identified two novel motifs (M1, M2) also enriched in the first 100 bp upstream of the TSS (p-value < $10^{-5}$, motif frequency > 20% higher than in other regions). M1 and M2 matched the transcription factor TFIIB (RNA polymerase II complex) downstream element (BREd), which has the consensus sequence G/A-T-T/G-A-T/G-G/T-T/G-T/G [58].

We argued that, if the expression of *de novo* human and chimpanzee genes was at least partly due to the co-option of genomic sequences as active promoters, we should observe a lower frequency of the relevant TFBS in the corresponding syntenic regions in macaque. This is exactly what we found for the five motifs mentioned earlier, whereas not differences existed for conserved genes (Figure 3C, S9 Fig). This was consistent with the gain of new transcription factor binding sites in the hominoid branches after the split from macaque. We also noted that the occurrence of transposable elements (Figure 3D) tended to decrease near the TSS of all gene classes except for endogenous retrovirus-derived long terminal repeats (LTRs), which overlap 13% of the proximal promoters of *de novo* genes compared to 5% of conserved genes. Further analyses indicated that LTRs tend to contribute CREB motifs (Figure 3E).

Transcription elongation is highly dependent on the presence of U1 small nuclear ribonucleoprotein recognition sites downstream of the TSS, whereas poly(A) sites (PAS) cause transcription termination [59]. As in standard multiexonic mRNAs, *de novo* genes showed enrichment of U1 sites and depletion of PAS downstream of the TSS. As U1 sites suppress the effect of PAS sites, we predicted that, if transcription elongation is restricted to hominoids, we should see an underrepresentation of U1 sites in the corresponding macaque syntenic regions, but not necessarily of PAS sites. We indeed observed this pattern in *de novo* genes, whereas no differences were detected for conserved genes (Figure 3F). This is consistent with the idea that the gain of U1 sites contributes to the stabilization of *de novo* genes.

**De novo originated proteins**

Most *de novo* genes were not annotated in the databases and their coding status was unclear. We analyzed two coding properties in *de novo* genes: ORF length and ORF score. The latter score was based on hexanucleotide frequencies in *bona fide* sets of coding and non-coding sequences (See Methods, [36]). The median length of the longest ORF in *de novo* genes was 52 amino acids, shorter than for proteins encoded by annotated coding RNAs (codRNA) with the same transcript length distribution as *de novo* genes and comparable to ORFs from similarly sampled intronic sequences (Figures 4A and 4B). In contrast, the coding score of the longest ORF was higher in *de novo* genes than in intronic ORFs (Wilcoxon test, p-value < $10^{-10}$) and comparable to the one for proteins shorter than 100 amino acids in the set of annotated protein-coding genes. Therefore, the results are consistent with an scenario in which a large fraction of *de novo* genes encode short proteins.

Next we searched for experimental evidence of proteins produced by *de novo* genes. We employed mass-spectrometry data from a recent study [60], limiting the searches to the same tissues used here for transcript assembly to increase specificity (testis, brain, heart and liver), and also searched in Proteomics DB [61]. We identified uniquely mapping peptides in 6 *de novo* genes, comprising 1 human- and 5 hominoid-specific genes (Table 1, Figure 4E). All of them were expressed in testis, and one of them was preferentially expressed in heart. In addition, we detected signatures of translation in 5 human- and 10 hominoid-specific *de novo* genes using available ribosome profiling sequencing data from human brain [62]. Overall 22 *de novo* genes, comprising 31 different transcripts, had evidence of translation.

Closer inspection of the genes with protein evidence showed that their size (median 76 amino acids) and coding potential (median 0.0414) were in line with values observed in *de novo* genes in general (Figures 4C and 4D). Two thirds of the ORFs were truncated in the syntenic region in macaque and none of them were detected in the syntenic region in mouse, consistent with a recent origination (S10 Fig). These genes also showed signatures of purifying selection (Table 2). This was assessed by calculating the fraction of nucleotide differences in different gene regions (introns, exons, ORF) with respect to the corresponding macaque syntenic genomic sequences. Sequences under selective constraints will have a lower number of substitutions than sequences evolving in a neutral manner. This is a conservative test, because selection is not expected to have acted in the macaque branch in *de novo* genes and positive selection may increase the number of substitutions. Despite this, we found that signatures of purifying selection could be clearly distinguished in ORFs with evidence of translation when compared to intronic regions, as it occurs in standard conserved coding sequences (Fisher-test, p-value < $10^{-5}$, Table 2). However, in general, the longest ORF in *de novo* genes did not show a significant decrease in the rate of divergence with respect to macaque, suggesting that the majority of these ORFs do not correspond to functional proteins or that positive selection is blurring the signal.

**DISCUSSION**

Here we performed a large-scale transcriptomics-based investigation of the emergence of new genes in hominoids. Our strategy was annotation-independent, which allowed us to recover many novel, non-annotated genes, and compare species for which the level of annotation varies greatly. The approach was entirely different from that employed in previous studies in which the initial datasets were composed of annotated protein coding genes in humans that lacked homologous proteins in other species [5,17,40,63]. We instead focused on new transcriptional events and subsequently analysed the coding potential of the transcripts. Therefore, the genes identified here are a more recent subset of transcriptional active loci or genes than those obtained previously.

We employed a polyadenylated RNA sequencing strategy that was based on a combination of high sequencing depth (average of 115 Million mapped reads per sample) and strand-specific sequencing. This, together with our annotation-blind strategy, resulted in the identification of 2,714 putative *de novo* genes expressing 5,398 transcripts, which is more than one order of magnitude the number found in previous studies [5,17,40,63]. The set of genes was obtained using a carefully chosen per base read coverage threshold, which allowed the full recovery of complete sequences while permitting the detection of transcripts expressed at low levels. Whereas our analysis was based on multiexonic genes we have to consider that many recently evolved genes may not have yet acquired the capacity to be spliced, as shown by several examples in *Drosophila* [7]. Therefore, there are probably many more *de novo* genes that those covered here. We found that *de novo* genes constituted about 4% of all expressed multiexonic genes in human and chimpanzee. Although this seems a very high fraction, it is consistent with previous transcriptomics-based studies in insects [64,65]. As these genes tended to be short and expressed at low levels, their associated transcriptional cost is relatively small. We showed that *de novo* human/hominoid genes show characteristic promoter and splicing signals and are expressed in a consistent manner across individuals.

The proportion of *de novo* genes with conserved genomic synteny in macaque was comparable to the proportion of conserved genes. Given the low number of nucleotide differences in neutrally evolving regions between the two species (~ 6%), we could reliably used syntenic alignments to examine transcription-related sequence features. We found an enrichment of transcription factor binding sites and U1snRNP motifs in the human and chimpanzee genes with respect to the corresponding genomic regions in macaque, consistent with the idea that the gain of regulatory motifs underlies *de novo* gene origination. This scenario had been proposed for the the formation of a new gene in mouse [66,67] but had not been tested until now at a genome-wide scale. Interestingly, in addition to general activators and polymerase II complex sites, we found an enrichment in RFX motifs. Although there are several members of the family that bind to similar sequences, many of the sites are probably recognized by RFX2, which is highly expressed in testis and has been involved in spermiogenesis [57].

Different studies dating the age of genes have found an excess of genes of very recent origin when compared to older gene classes [65,64]. This suggests that many young genes are subsequently lost, which is consistent with the relatively constant number of genes observed in a taxon. We found that genes with evidence of translation had significant signatures of purifying selection, indicating that at least some of them are functional. However, the signature was very weak for *de novo* genes in general, which would be consistent with a scenario in which many of these genes are dispensable. In contrast, studies in *Drosophila* indicate that directional selection determines the fate of *de novo* genes from the very early stages [7]. Although we focused on possible coding functions, some of these genes may have non-coding functions. This is specially relevant in the case of antisense transcripts which can potentially influence the expression of the transcript in the opposite orientation [68]. It is important to consider that the annotations alone may not suffice to differentiate between coding and non-coding transcripts as many annotated lncRNAs may translate short peptides according to ribosome profiling data [33,35,69]. The coding score of *de novo* genes was clearly non-random. One possible explanation is that natural selection rapidly eliminate transcripts that produce toxic peptides [70], as one can expect such peptides to often have unusual amino acid compositions.

Here we detected 20 putative new human proteins using ribosome profiling from brain [71]. Considering that the expression of most *de novo* genes was restricted to testis, for which no ribosome profiling data has yet been published, we expect this number to increase substantially in the future. Mass-spectrometry has important limitations for the detection of short peptides [72]., but we could nevertheless detect 8 putative proteins, mostly from testis. Our results indicate that expression of new loci in the genome takes place at a very high rate and is probably mediated by random mutations that generate new active promoters. The new genes form the substrate for the evolution of new functions and species-specific adaptations.

**MATERIALS AND METHODS**

**Ethics statement**

Chimpanzee and macaque samples were obtained from the Primate Bio-Bank of the Biomedical Primate Research Center (BPRC). BPRC offers state-of-the-art animal facilities (AAALAC accredited) and is fully compliant with regulations on the use of non-human primates for medical research. BPRC's Primate Tissue Bank is one of the biggest non-human primate banks in Europe and it is involved in the framework of the EuprimNet Bio-Bank (www.euprim-net.eu). The EUPRIM-Net Bio-Bank is conducted and supervised by the scientific government board along all lines of EU regulations and in harmonisation with Directive 2010/63/EU on the Protection of Animals Used for Scientific Purposes. The animals used for tissue collection in all cases are diagnosed with cause of death other than their participation in this study and without any relation to the tissues used.

**Library preparation and strand-specific polyA+ RNA-Seq protocol**

Human and mouse total RNA was purchased from Amsbio. Chimpanzee and macaque total RNA was extracted using a miRNeasy Mini kit from tissue samples obtained at the Biomedical Primate Research Centre (BPRC, Netherlands). Mouse samples were from a pool of 3 males and 3 females (Balb/C strain).

Libraries were prepared using the TruSeq Stranded mRNA Sample Prep Kit v2 according to the manufacturer's protocol. PolyA+ RNA was purified from 250-500 mg of total RNA using streptavidin-coated magnetic beads (AMPure XP) and subsequently fragmented to ~300 bp. cDNA was synthesized using reverse transcriptase (SuperScript II, Invitrogen) and random primers. We did not add reverse transcriptase to one of the human testis replicate samples to use it as a control of DNA contamination (RT-). The strand-specific RNA-Seq library preparation was based on the incorporation of dUTP in place of dTTP in the second strand of the cDNA. Double-stranded DNA was further used for library preparation. Such dsDNA was subjected to A-tailing and ligation of the barcoded Truseq adapters. Library amplification was performed by PCR on the size selected fragments using the primer cocktail supplied in the kit. Sequencing was done with a Illumina HiSeq 2000 sequencer in a paired-end design (2x100 nt) according to the manufacturer's instructions. Library preparation and sequencing were done at the Genomics Unit of the Center for Regulatory Genomics (CRG, Barcelona, Spain).

**RNA-Seq datasets**

The polyA+ RNA-Seq included 96 sequencing datasets for 9 different species: 43 strand-specific paired end data (~3 billion reads) and 53 single read data (~3.2 billion reads). The strand-specific data was employed for the assembly of reference transcripts for human, chimpanzee, macaque and mouse (Fig. 1 for a summary of results). For comparative purposes, we used the same tissues and number of biological samples for human and chimpanzee (liver, heart, brain and testis; two biological replicates per tissue). For macaque and mouse, the outgroup species, we added available strand-specific RNA-Seq data from other tissues: adipose, skeletal muscle for macaque [40], ovary and placenta for mouse [41,47]. The single read data corresponded to 5 primate species (human, chimpanzee, gorilla, orangutan, macaque) and 4 additional vertebrates (mouse, chicken, platypus and opossum) in 6 different tissues (brain, cerebellum, heart, kidney, liver and testis) [44]. Although these experiments were based on single reads and had lower coverage than the strand-specific RNA-Seq data, they were useful to increase the number of species with expression data for sequence similarity searches. More information about the samples can be found in supplementary file 2.

**Read mapping and transcriptome assembly**

RNA-Seq sequencing reads underwent quality filtering using Condetri (v.2.2) [73] with the following settings (-hq=30 –lq=10). Adapters were trimmed from filtered reads if at least 5 nucleotides of the adaptor sequence matched the end of each read. In all experiments, reads below 50 nucleotides or with only one member of the pair were not considered. We retrieved genome sequences and gene annotations from Ensembl v. 75 [74]. We aligned the reads to the correspondent reference species genome with Tophat (v. 2.0.8) [75] with parameters –N 3, -a 5 and –m 1, and including the correspondent parameters for paired-end and strand-specific reads whenever necessary. Multiple mapping to several locations in the genome was allowed unless otherwise stated.

We performed gene and transcript assembly with Cufflinks (v 2.2.0) [76] for each individual sample. Per-base read coverage and FPKM (fragments per kilobase of transcript per million mapped fragments) values were calculated for each transcript and gene as described by Trapnell et al. (2010). We only considered assembled transcripts that met the following requirements: a) the transcript was covered by at least 4 reads, b) Abundance was higher than 1% of the most abundant isoform of the gene and, c) <20% of reads were mapped to multiple locations in the genome.

Subsequently, we used Cuffmerge [76] to build a single set of assembled transcripts per each species, separately for the strand-specific and the single read based RNA-Seq experiments. We compared our set of assembled transcripts with gene annotation files from Ensembl (gtf format, v.75) with Cuffcompare [76] to identify transcripts corresponding to annotated genes. This included the categories '=' (complete match), 'c' (contained), 'j' (novel isoform), "e" and "o" (other exonic overlaps in the same strand). Genes for which none of the assembled transcripts matched an annotated gene were labeled novel. In human, 82% and 44.5% of the total annotated protein-coding and non-coding genes (lincRNA, antisense and processed transcripts), respectively, were recovered.

Additionally, we run Trinity [42], which reconstructs transcripts in the absence of a reference genome, with all unmapped reads in each species (read length >= 75 nucleotides). Before running Trinity, unmapped reads were normalized by median using Khmer (parameters –C 20, -k 20, -N 4). This allowed the recovery of any transcripts falling into non-assembled parts of the genome. We selected transcripts with a minimum size of 300 nucleotides.

We obtained a set of reference transcripts from the strand-specific RNA-Seq data using a per-nucleotide read coverage >= 5. This choice was based on the relationship between read coverage and percentage of fully reconstructed annotated coding regions (CDS, longest one per gene) for the subset of genes mapping to annotated protein coding genes (Ensembl v.75), using only the categories '=' and 'c' in Cuffcompare (18,694 protein-coding genes). For values higher than 5 there was no substantial increase in the number of fully reconstructed CDS (coverage >= 5: 87.8%; coverage >= 10: 88.5%; coverage >= 20: 89.4%). The selection was based on coding regions and not complete transcripts because of the known existence of alternative transcription start sites in many annotated transcripts causing uncertainty in the latter parameter [77]. Very similar results were obtained for CDS shorter than 500 nucleotides or genes with only one annotated CDS, indicating that protein length or gene complexity has little effect on the suitability of this threshold.

Transcript assembly with the RT- control (see above) resulted in 22,803 different sequences that presumably corresponded to genomic DNA contamination, resulting from regions resistant to DNAse treatment. Except for the reverse transcriptase, all other reagents were added in the same concentration as in the other samples. Therefore, the number of contaminant fragments must be considered an upper boundary, as in a normal RNA-Seq experiment these fragments are probably sequenced much less efficiency as they have to compete with the genuine RT products. The sequences obtained in the RT- control did not contain any introns and the majority of them were shorter than 300 nucleotides (98.58%).

**Genomic comparisons**

Reference transcripts were classified into three categories depending on their location with respect to transcripts from other genes: a) Intergenic: Transcripts that did not overlap any other assembled loci. b) Overlappping Intronic: Transcripts located within introns of other assembled genes in the opposite strand. c) Overlapping antisense: Transcripts partially or completely overlapping exons from other assembled genes in the opposite strand.

We downloaded long interspersed element (LINE), short interspersed element (SINE) and long terminal repeat (LTR) annotations in the human and chimpanzee genomes from RepeatMasker (same genome versions than in Ensembl v.75) [78]. We used BEDTools [79] to identify any overlap between transcripts and/or genomic elements.

We downloaded pairwise syntenic genomic alignments of human against chimpanzee, macaque and mouse; and of chimpanzee against human, macaque and mouse, aligned by the blastz alignment [80], from UCSC. We developed an in-house Python script to recover syntenic regions corresponding to a given human or chimpanzee transcript or to regions upstream and downstream of a human or chimpanzee transcription start site (TSS).

We scanned the human and chimpanzee genomes to identify transcripts with bidirectional promoters. We recovered any antisense pairs in which the distance between the two TSSs was < 1 kb). For annotated genes in general the frequency was 29.81%, compared to 20% for *de novo* genes. This was significantly higher than that expected by chance (Binomial Test, p-value << $10^{-5}$, 20% versus 5.31%).

**Identification of *de novo* genes**

We developed a pipeline to identify *de novo* genes in human and chimpanzee based on the lack of homologues in other species. We first selected multiexonic transcripts from the reference transcriptome assemblies. Then, we performed exhaustive sequence similarity searches against sequences from other species with the BLAST suite of programs and, subsequently, searched for overlapping transcripts in genomic syntenic regions.

Sequence similarity searches, using reference human or chimpanzee transcripts as query, were performed against the complete transcriptome assemblies from the nine different vertebrate species, gene annotations from Ensembl v.75 for the same species and the EST and non-redundant protein "nr" [81] NCBI databases. We employed both BLASTN and TBLASTX programs [43], with an E-value threshold of $10^{-4}$. All BLAST searches were performed with the filter of low-complexity regions activated and so we discarded any transcripts in which self-hits were not reported. Species-specific genes were those for which no transcripts (or transcripts from any paralogs) had sequence similarity hits to transcripts in any other species. We identified 634 human-specific genes (1,029 transcripts) and 780 chimpanzee-specific genes (1,307 transcripts). In the case of hominoid-specific genes we allowed for hits to gorilla and orangutan in addition to human and chimpanzee. The pipeline yielded 1,300 hominoid-specific genes (3,062 transcripts). About one third of them (221 genes and 1,016 transcripts) were reference transcripts in both species (multiexonic, coverage >= 5) and the rest were identified via the complete transcriptome assemblies, EST and/or nr databases. Because not all of them were detected as reference transcripts in both species the number of hominoid-specific genes is different for human and chimpanzee (604 and 916, respectively).

For synteny-based identificaton of homologues we took advantage of the existing pairwise syntenic genomic alignments from UCSC. We employed data from human, chimpanzee, macaque and mouse. If two transcripts overlapped (>= 1bp) in the syntenic region we considered it as evidence of homology. We reclassified the *de novo* genes accordingly.

**Tissue gene expression**

We analyzed the patterns of tissue expression in assembled transcripts, considering a transcript as expressed in one tissue if FPKM > 0. We measured the number of tissue-restricted transcripts using a previously proposed metric [82]:

$$\tau = \frac{\sum_{i=n}^{i=1}(1-x_i)}{N-1}$$

Where N is the number of tissues and $x_i$ is the FPKM expression value of the transcript in the sample normalized by the maximum expression value over all tissues. We classified cases with a τ > 0.85 as preferentially expressed in one tissue or tissue-restricted.

For *de novo* genes annotated in Ensembl v.75 we obtained expression data from the GTEx project, which comprises a large number of human tissue samples. We used this data to calculate the number of genes showing tissue-restricted expression as well as the number of testis samples with detectable expression of a given gene.

**Motif analysis**

We searched for significantly overrepresented motifs in *de novo* and conserved genes using computational approaches. We employed sequences spanning from 300 bp upstream to 300 bp downstream of the transcription start site (TSS). Redundant TSS positions were only considered once. With PEAKS [54] we identified three TRANSFAC motifs [83] enriched in *de novo* genes, corresponding to CREB, JUN, RFX (p-value < $10^{-5}$ and minimum of 30 motif occurrences). With HOMER [55] we identified two novel motifs (M1, M2) enriched in *de novo* genes in the first 100 bp upstream of the TSS (p-value < $10^{-5}$ and enrichment > 20% when compared to other regions). M1 and M2 matched the transcription factor TFIIB (RNA polymerase II complex) downstream element (BREd), which has the consensus sequence G/A-T-T/G/A-T/G-G/T-T/G-T/G [58].

For graphical representation of the results, we computed the relative motif density in 100 bp windows upstream and downstream of the TSS in human and chimpanzee, and the corresponding genomic syntenic regions in macaque and mouse. We used MEME [84] to scan the sequences for the occurrence of motifs (matches to weight matrices with a p-value < $10^{-5}$). The average number of motif occurrences (motif density) was normalized to values between 0 and 1, where 1 corresponded to the highest density of a given motif in a sequence window.

It has been previously proposed that new genes tend to gain new U1 sites and lose PAS sites as they become more mature [85]. We used MEME with the same parameters as described above to search for U1 (U1 snRNP 5' splice site consensus motif) and PAS (poly-adenylation signals) sites 500 bp sequences upstream and downstream of the TSS (see Sampledata.xlsx for weight matrices). PAS motifs found < 500bp downstream of a U1 site were not considered since the PAS effect is abolished by snRNPs bound to these U1 motifs at such distances.

**Coding score**

We defined an open reading frame (ORF) in a transcript as any sequence starting with an ATG codon and finishing at a stop codon (TAA, TAG or TGA). In addition we require it to be at least 75 nucleotides long (24 amino acids), which is the the size of the smallest complete human polypeptide found in genetic screen studies [86].

In each ORF we computed a coding score based on hexamer frequencies in *bona fide* coding and non-coding sequences [36]. Specifically, we first computed one coding score (CS) per nucleotide hexamer:

$$CS_{hexamer(i)} = \log\left(\frac{freq_{coding}(hexamer(i))}{freq_{non-coding}(hexamer(i))}\right)$$

The coding hexamer frequencies were obtained from all human transcripts encoding experimentally validated proteins. The non-coding hexamer frequencies were calculated using the longest ORF in intronic regions, which were selected randomly from expressed protein-coding genes. The hexamer frequencies were computed separately for ORFs with different lengths to account for any possible length-related biases (24-39, 40-59, >60 amino acids). Next, we used the following statistic to measure the coding score of an ORF:

$$CS_{ORF} = \frac{\sum_{i=n}^{i=1} CS_{hexamer(i)}}{n}$$

where i is each sequence hexamer in the ORF, and n the number of hexamers considered.

The hexamers were calculated in steps of 3 nucleotides in frame (dicodons). We did not consider the initial hexamers containing a Methionine or the last hexamers containing a STOP codon. Given that all ORFs were at least 75 nucleotides long the minimum value for n was 22.

In coding RNAs (CodRNA all) the annotated ORF was selected for further analysis. In addition, to account for possible due to biases in transcript length, we randomly selected a subset of protein-coding transcripts with the same transcript length distribution as transcripts from *de novo* genes (CodRNA short). In sequences with no annotated coding sequence (introns and transcripts from *de novo* genes), we chose the longest ORF considering all three possible frames. The only exception was when the longest ORF in another frame had a higher coding score than expected for non-coding sequences (0.0448 if ORF < 40 aa; 0.0314 if 60 aa > length ORF >= 40 aa; 0.0346 if length ORF >= 60 aa; p-value < 0.05) or it was longer than expected for non-coding sequences (>= 134 aa, p-value < 0.05), in which case we selected this other ORF (in 3.4% of the cases).

**Ribosome profiling data**

We downloaded data from ribosome profiling experiments in human brain tissue [62]. Ribosome profiling reads were filtered as described previously [36]. We then used Bowtie2 [87] to map the reads to the human assembled transcripts with no mismatches. We considered each strand independently since the RNA-Seq data was strand-specific. RNA-Seq reads from the same experiment were also mapped to *de novo* transcripts to determine how many of them were expressed (FPKM > 0). Because of the low detectability of ribosome association at low FPKM expression values [36], two ribosome profiling reads mapping to a predicted ORF were deemed sufficient for the signal to be reported.

**Mass spectrometry data**

We used available mass-spectrometry data from human frontal cortex, liver, heart and testis [60,61] to identify any putative peptides produced by *de novo* genes. Mass-spectrometry data was searched using the Proteome Discoverer software v.1.4.1.14 (Thermo Fisher Scientific, United States) using MASCOT v2.5 [88] as search engine. The database used contained the human entries in SwissProt [89], the most common contaminants and putative peptides derived from the translation of transcripts from *de novo* genes. Carbamidomethylation for cysteines was set as fixed modification whereas acetylation in protein N-terminal and oxidation of methionine were set as variable modifications. Peptide tolerance was 7 ppm in MS and 20mmu in MS/MS mode, maximum number of missed cleavages was set at 3. The Percolator [90] algorithm implemented in the Proteome Discoverer software was used to estimate the qvalue and only peptides with qvalue < 0.01 and rank = 1 were considered as positive identifications. Lastly, we only considered unique peptides matching to young transcripts by using BLAST with short query parameters to search the candidate peptides against all predicted ORFs in assembled transcripts. Additionally, we searched for evidences of peptides assigned to our set of *de novo* annotated genes in Proteomics DB [61]. We found 6 *de novo* genes with proteomics evidence, two of them were annotated in Ensembl as lncRNAs and expressed in ≥55 testis samples from GTEx. Details of the results can be found in supplementary file 2.

**Statistical data analyses and plots**

The analysis of the data, including generation of plots and statistical test, was done using R [91].

**SUPPORTING INFORMATION**

Supplementary file 1 contains the supplementary tables and figures (Table S1 and Figures S1-S10). Supplementary file 2 contains information of the RNA-Seq datasets, proteomics hits and motif weight matrices. Annotation files of *de novo* genes identified here are available at evolutionarygenomics.imim.es (Publications, Datasets, ptr_denovo.gtf for chimpanzee gene coordinates and hsa_denovo.gtf for human gene coordinates).

**FIGURE LEGENDS**

**Figure 1. Global properties of assembled transcriptomes. a)** Percentage of annotated and novel genes and transcripts using strand-specific deep polyA+ RNA sequencing. Classification is based on the comparison to reference gene annotations in Ensembl v.75. 70.65 and 87.77% of annotated genes in human and mouse are classified as protein-coding. Detailed statistics on the total and per-tissue number of assembled genes and transcripts is available in the supplementary material. **b)** Cumulative density of nucleotide length in annotated and novel assembled transcripts. **c)** Cumulative density of expression values in logarithmic scale in annotated and novel assembled transcripts. Expression is measured in fragments per kilobase per million mapped reads (FPKM) values, selecting the maximum value across all samples.

**Figure 2. Identification and characterization of *de novo* genes in human and chimpanzee. a)** Simplified phylogenetic tree indicating the nine species considered in this study. In all species we had RNA-Seq data from several tissues. Chimpanzee, human, macaque and mouse were the species for which we performed strand-specific deep polyA+ RNA sequencing. We indicate the branches in which *de novo* genes were defined, together with the number of genes. **b)** Categories of transcripts in *de novo* genes based on genomic locations. Intergenic, transcripts that are not overlapping any other gene; Overlapping antisense, transcripts that overlap exons from other genes in the opposite strand; Overlapping intronic, transcripts that overlap introns from other genes in the opposite strand, with no exonic overlap. **c)** Classification of *de novo* genes based on existing evidence in databases. Annotated; genes classified as annotated in Ensembl v.75; EST/nr; non-annotated genes with BLAST hits ($10^{-4}$) to expressed sequence tags (EST) and/or non-redundant protein (nr) sequences in the same species. Novel; rest of genes. **d)** Patterns of gene expression in four tissues. Brain refers to frontal cortex. Transcripts with FPKM > 0 in a tissue are considered as expressed in that tissue. In red boxes, fraction of transcripts whose expression is restricted to that tissue (t > 0.85, see Methods). Chimp conserved, transcripts assembled in chimpanzee not classified as *de novo*. Human conserved, transcripts assembled in human not classified as *de novo*. **e)** Number of testis GTEx samples with expression of *de novo* and conserved genes. We considered all annotated genes with FPKM > 0 in at least one testis sample. Conserved, genes sampled from the total pool of annotated genes analyzed in GTEx with the same distribution of FPKM values than in annotated *de novo* genes (n=200).

**Figure 3. Recent signatures of transcription in *de novo* genes. a)** Overrepresented transcription factor binding sites (TFBS) in the region -100 to 0 with respect to the transcription start site (TSS) in *de novo* genes. The region from -300 to +300 with respect to the TSS was analysed (n=3,875). Color code relates to normalized values (highest value is yellow). **b)** Fine-grained motif density 200bp upstream of the TSS is shown. **c)** Comparison of motif density in genomic syntenic regions in macaque for *de novo* transcripts (n=3,116) and conserved transcripts (n=4,323, randomly taken human and chimpanzee annotated transcripts not classified as *de novo*). Significant differences between human/chimpanzee and macaque are indicated; Fisher-test; *, p-value < 0.05; **, p-value < 0.01. **c)** Comparison of motif density in promoters with and without long terminal repeat (LTR) in the region -500 to 0 with respect to the TSS. Significant differences in motif density in the -100 bp window are indicated. **d)** Density of the main human transposable elements (TE) families around the TSS of *de novo* and conserved transcripts. Regions -3 kB to +3 kB with respect to the TSS were analysed. LTR frequency is higher in the region -100 to +100 in de novo genes when compared to conserved genes (Fisher-test p-value < $10^{-18}$). **e)** Signatures of transcription elongation in *de novo* and conserved genes. Density of U1 and PAS motifs in the 500bp region upstream and downstream of the TSS. Comparison of U1 and PAS motif density in genomic syntenic regions in macaque for *de novo* transcripts (n=3,116) and conserved transcripts (n=4,323). There is an increase of U1 motifs in *de novo* transcripts when compared to macaque (indicated by a black arrow, Fisher-test, p-value=0.016 for the region +100 to +200).

**Figure 4. Coding potential of *de novo* genes. a-d)** ORF length and coding score for ORFs in different sequence types. *De novo* gene, longest ORF in *de novo* transcripts (n=1,933). CodRNA (all), annotated coding sequences from Ensembl v.75 (n=8,462). CodRNA (short), annotated coding sequences sampled as to have the same transcript length distribution as *de novo* transcripts (n=1,952). Intron, longest ORF in intronic sequences from annotated genes sampled as to have the same transcript length distribution as *novo* transcripts (n=5,000); Proteogenomics – ORFs in *de novo* transcripts with peptide evidence by mass-spectrometry; Ribosome profiling – ORFs in *de novo* transcripts with ribosome association evidence in brain. **e)** Example of hominoid-specific *de novo* gene with evidence of protein expression from proteogenomics, with RNA-Seq read profiles in two human samples. **(f)** Example of hominoid-specific *de novo* gene with RNA-Seq and ribosome profiling read profiles. Predicted coding sequences are highlighted with red boxes and the putative encoded protein sequences displayed.

**ACKNOWLEDGEMENTS**

We received funding from Ministerio de Economía y Competitividad (BFU2012-36820 co-funded by FEDER to M.M.A.), the Secretariat of Universities and Research-Government of Catalonia (2014SGR1121) and Institució Catalana de Recerca i Estudis Avançats (ICREA contract to M.M.A. and T.M.-B.). The CRG/UPF Proteomics Unit is part of the "Plataforma de Recursos Biomoleculares y Bioinformáticos (ProteoRed)" supported by grant PT13/0001 of Instituto de Salud Carlos III (ISCIII). T.M.-B. Was supported by EMBO YIP 2014. We are grateful for many discussions with many colleagues during the progress of this work.**AUTHOR CONTRIBUTIONS**

J.R.-O. and M.M.A. conceived the study, J.R.-O. performed the data analysis, I.K and R.B. provided samples, J.H.-R. and T.M.-B. performed RNA extractions, E.S and C.C performed the proteomics analysis, J.R.-O. and M.M.A. interpreted the data and wrote the manuscript.

**AUTHOR INFORMATION**

Sequencing data is deposited in the Gene Expression Omnibus under accession number GSE69241.

**REFERENCES**

1. Basu K, Graham LA, Campbell RL, Davies PL (2015) Flies expand the repertoire of protein structures that bind ice. Proc Natl Acad Sci U S A 112: 737–742. doi:10.1073/pnas.1422272112.

2. Bosch TCG (2014) Rethinking the role of immunity: lessons from Hydra. Trends Immunol 35: 495–502. doi:10.1016/j.it.2014.07.008.

3. Ohno S (1970) Evolution by gene duplication. Unwin A, editor Springer-Verlag. 160 p.

4. Levine MT, Jones CD, Kern AD, Lindfors HA, Begun DJ (2006) Novel genes derived from noncoding DNA in Drosophila melanogaster are frequently X-linked and exhibit testis-biased expression. Proc Natl Acad Sci U S A 103: 9935–9939. doi:10.1073/pnas.0509809103.

5. Toll-Riera M, Bosch N, Bellora N, Castelo R, Armengol L, et al. (2009) Origin of primate orphan genes: a comparative genomics approach. Mol Biol Evol 26: 603–612. doi:10.1093/molbev/msn281.

6. Arendsee ZW, Li L, Wurtele ES (2014) Coming of age: orphan genes in plants. Trends Plant Sci 19: 698–708. doi:10.1016/j.tplants.2014.07.003.

7. Zhao L, Saelao P, Jones CD, Begun DJ (2014) Origin and spread of de novo genes in Drosophila melanogaster populations. Science 343: 769–772. doi:10.1126/science.1248286.

8. Jacob F (1977) Evolution and tinkering. Science 196: 1161–1166.

9. Kaessmann H (2010) Origins, evolution, and phenotypic impact of new genes. Genome Res 20: 1313–1326. doi:10.1101/gr.101386.109.

10. Soukup SW (1974) Evolution by gene duplication. S. Ohno. Springer-Verlag, New York. 1970. 160 pp. Teratology 9: 250–251. doi:10.1002/tera.1420090224.

11. Cai J, Zhao R, Jiang H, Wang W (2008) De novo origination of a new protein-coding gene in Saccharomyces cerevisiae. Genetics 179: 487–496. doi:10.1534/genetics.107.084491.

12. Ohno S (1984) Birth of a unique enzyme from an alternative reading frame of the preexisted, internally repetitious coding sequence. Proc Natl Acad Sci U S A 81: 2421–2425.

13. Chen S, Zhang YE, Long M (2010) New genes in Drosophila quickly become essential. Science 330: 1682–1685.

14. Reinhardt JA, Wanjiru BM, Brant AT, Saelao P, Begun DJ, et al. (2013) De novo ORFs in Drosophila are important to organismal fitness and evolved rapidly from previously non-coding sequences. PLoS Genet 9: e1003860. doi:10.1371/journal.pgen.1003860.

15. Knowles DG, Mclysaght A (2009) Recent de novo origin of human protein-coding genes. Genome Res 19: 1752–1759. doi:10.1101/gr.095026.109.

16. Wu D-D, Irwin DM, Zhang Y-P (2011) De novo origin of human protein-coding genes. PLoS Genet 7: e1002379. doi:10.1371/journal.pgen.1002379.

17. Chen J-Y, Shen QS, Zhou W-Z, Peng J, He BZ, et al. (2015) Emergence, Retention and Selection: A Trilogy of Origination for Functional De Novo Proteins from Ancestral LncRNAs in Primates. PLoS Genet 11: e1005391. doi:10.1371/journal.pgen.1005391.19

**Table 1**

| Detection Technique | Assembly gene ID | Assembly transcript ID | Age[c] | Tissue[d] | Protein length | Annotation[e] |
|---|---|---|---|---|---|---|
| Proteogenomics[a] | XLOC_175402 | hsa_00362506 | Hominoid | Heart | 36 | LncRNA (ENSG00000223485) |
| | XLOC_068697 | hsa_00142705 | Hominoid | Testis | 37 | Novel |
| | XLOC_085716 | hsa_00181285 | Hominoid | Testis | 64 | Novel |
| | XLOC_088783 | hsa_00187116, hsa_00187117, hsa_00187118 | Hominoid | Testis | 148, 136, 61 | LncRNA (ENSG00000263417) |
| | XLOC_105288 | hsa_00223807 | Hominoid | Testis | 199 | Novel |
| | XLOC_196865 | hsa_00404039 | Human | Testis | 49 | Novel |
| Ribosome profiling[b] | XLOC_002919 | hsa_00006742, hsa_00006743, hsa_00006744 | Hominoid | Brain, Heart | 68, 64, 58 | Novel |
| | XLOC_031861 | hsa_00068400 | Human | Brain | 58 | LncRNA (ENSG00000273409) |
| | XLOC_042102 | hsa_00090118 | Hominoid | Brain | 90 | LncRNA (ENSG00000257061) |
| | XLOC_050821 | hsa_00107269 | Human | Brain | 56 | Novel |
| | XLOC_057303 | hsa_00119633 | Hominoid | Testis, Brain | 52 | Novel |
| | XLOC_073846 | hsa_00154236 | Hominoid | Brain | 54 | Novel |
| | XLOC_082421 | hsa_00173626, hsa_00173627 | Hominoid | All 4 tissues | 95, 95 | LncRNA (ENSG00000265666) |
| | XLOC_085590 | hsa_00181107, hsa_00181108 | Hominoid | Brain, Testis | 89, 83 | Novel |
| | XLOC_104066 | hsa_00221170 | Hominoid | Brain | 68 | Novel |
| | XLOC_106910 | hsa_00227119 | Human | Brain | 36 | LncRNA (ENSG00000228999) |
| | XLOC_152506 | hsa_00317537 | Hominoid | Brain | 53 | LncRNA (ENSG00000251423) |
| | XLOC_160844 | hsa_00333276, hsa_00333277 | Hominoid | Brain | 65, 65 | Novel |
| | XLOC_168602 | hsa_00348960 | Hominoid | Brain | 29 | LncRNA (ENSG00000228408) |
| | XLOC_184660 | hsa_00380291 | Human | Brain | 101 | LncRNA (ENSG00000236197) |
| | XLOC_195038 | hsa_00400469 | Human | Brain | 42 | novel |

**Table 1. Human *de novo* genes with evidence of protein translation.** [a] Proteogenomics, detection is based on the identification of mass spectrometry peptides with a unique match to an ORF and corrected p-value (q-value) < 0.01 (brain, heart, liver and testis data from Kim et al., 2014). [b] Ribosome profiling, detection is based on the presence of ribosome profiling reads overlapping the ORF (brain data from Gónzalez et al., 2014). [c]Age refers to whether the gene is human-specific or hominoid-specific. [d]The tissue with preferential expression is indicated, using the RNA-Seq data generated here for human brain, heart, liver and testis. [e]Annotation refers to the classification of the transcripts as novel or annotated in Ensembl v.75.

**Table 2**

| Dataset | Transcript | | | Introns[c] |
| --- | --- | --- | --- | --- |
| | Complete exons | ORF[b] | Rest exonic sequence | |
| 1. Species-specific *de novo* transcripts | 58.74 *** | 61.92 | 57.75 | 61.47 |
| 2. Hominoid-specific *de novo* transcripts | 59.11 *** | 60.76 | 58.67 | 60.42 |
| 3. *De novo* transcripts with protein evidence | 48.71 *** | 43.51 ** | 49.91 | 59.65 |
| 4. Conserved transcripts | 43.85 *** | 29.39 *** | 62.47 | 60.14 |

**Table 2. Divergence with macaque in different gene regions.** Number of nucleotides differences per Kb in macaque syntenic regions with respect to human or chimpanzee sequences. Protein evidence is from proteomics or ribosome profiling (Table 1). We did not consider regions with gaps or transcripts with partial synteny. [b]ORF: refers to the longest ORF in transcripts from *de novo* genes, or with protein evidence/annotated in the rest. [c]Introns: sampled intronic regions of size 500 pb from the same set of transcripts. Conserved: conserved human protein coding transcripts annotated in Ensembl v.75. We tested for differences between complete exons and introns, and ORF and introns with the Fisher test: *p-value < 0.05, **p-value<0.005, ***p-value < $10^{-5}$.

**Figure 1**

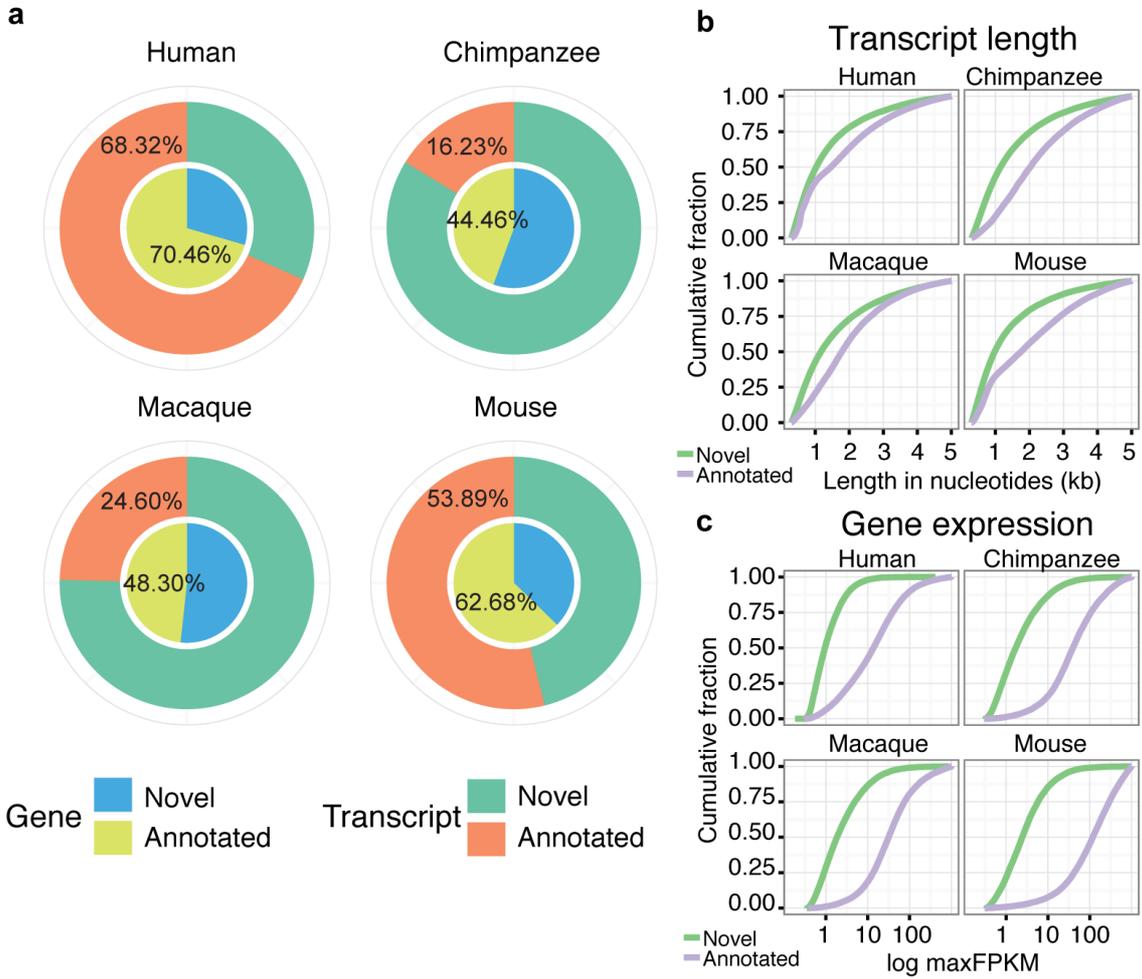

**Figure 2**

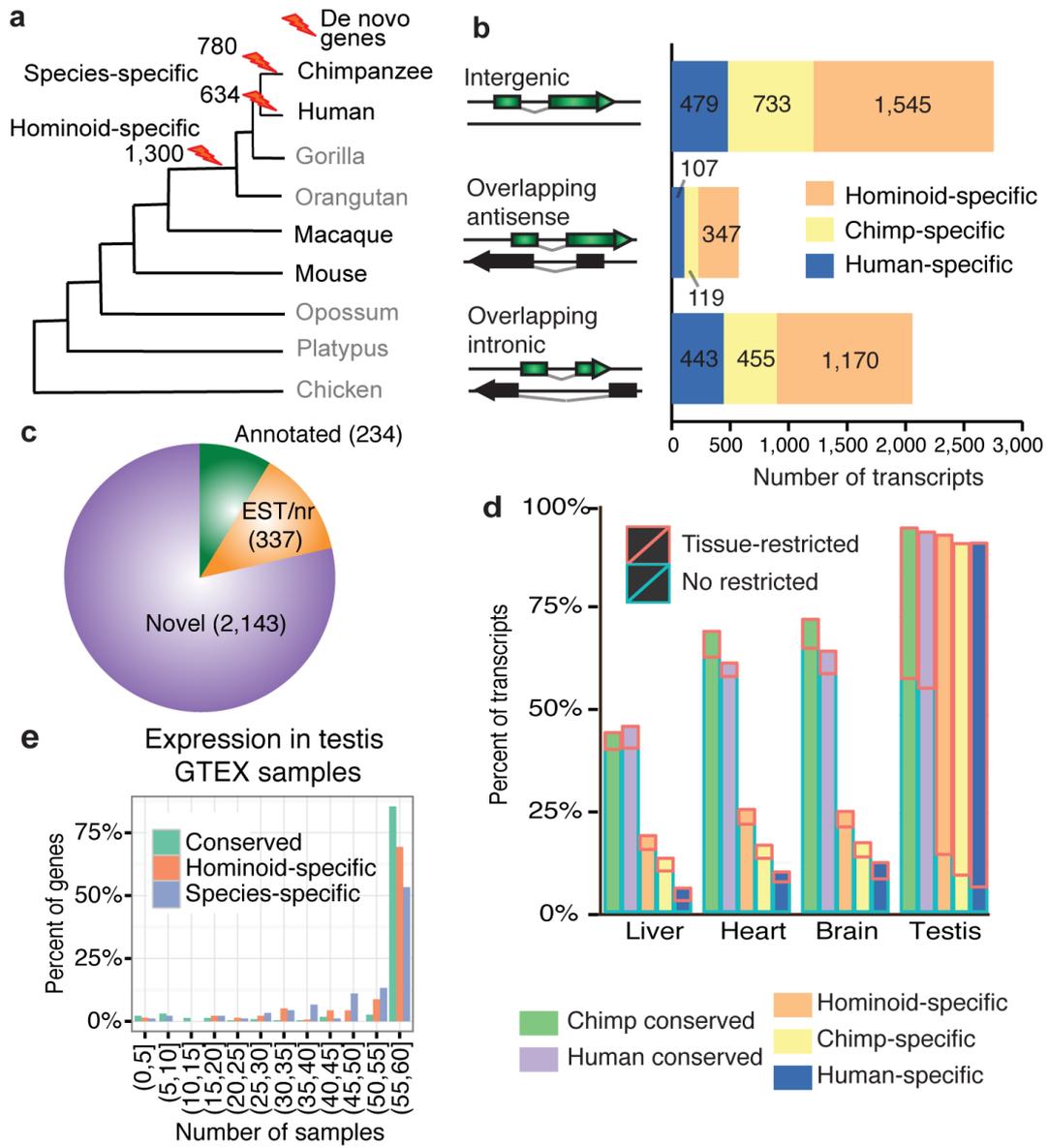

**Figure 3**

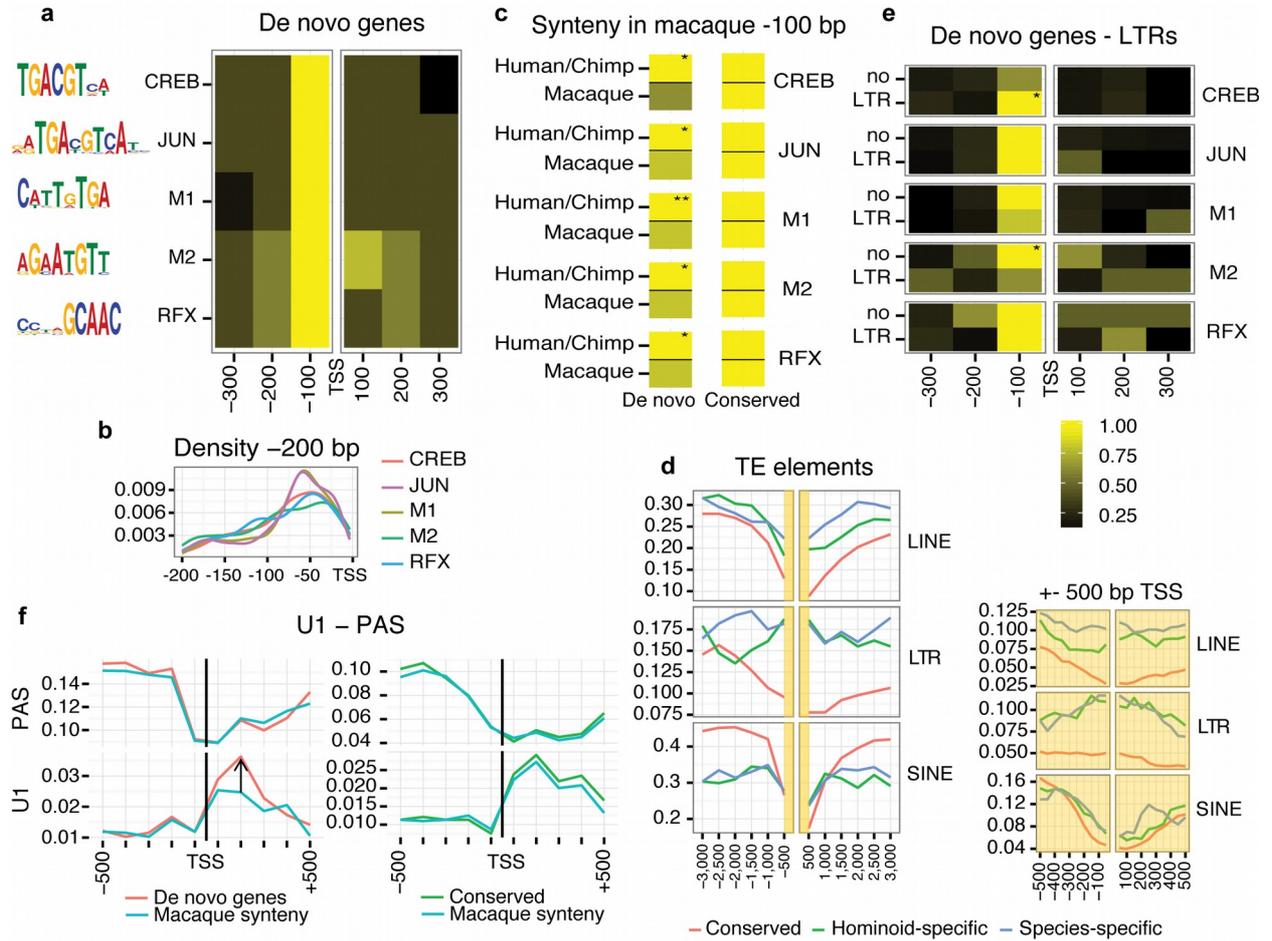

**Figure 4**

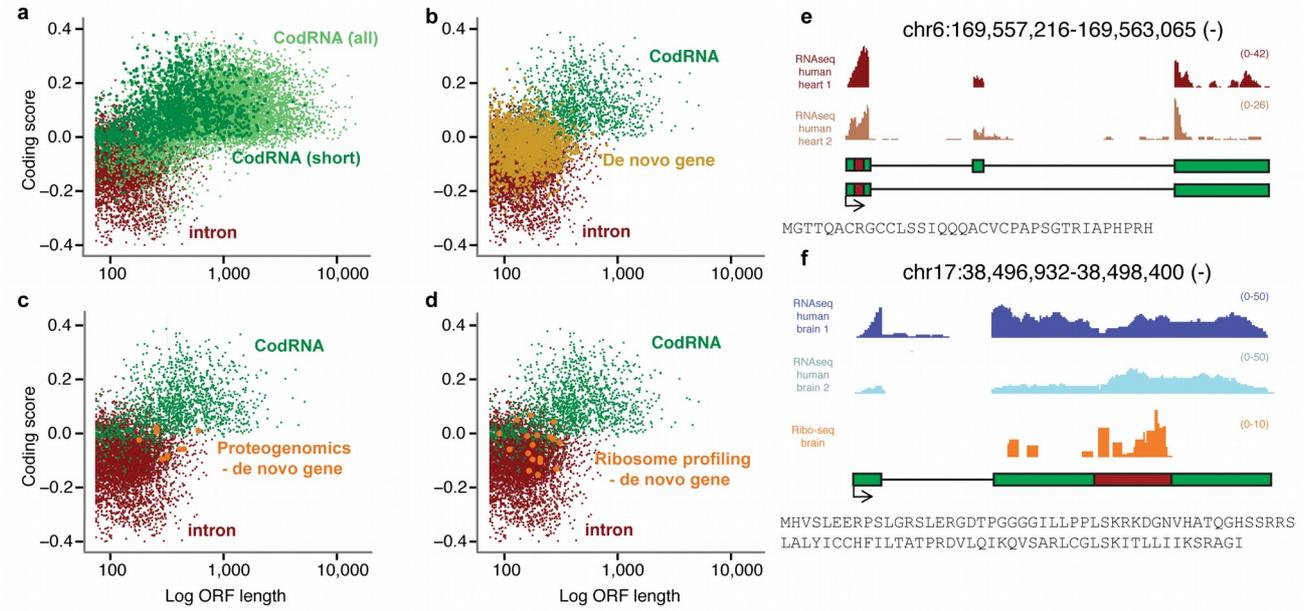